

\def\bbbc{{\mathchoice {\setbox0=\hbox{$\displaystyle\rm C$}\hbox{\hbox
to0pt{\kern0.4\wd0\vrule height0.9\ht0\hss}\box0}}
{\setbox0=\hbox{$\textstyle\rm C$}\hbox{\hbox
to0pt{\kern0.4\wd0\vrule height0.9\ht0\hss}\box0}}
{\setbox0=\hbox{$\scriptstyle\rm C$}\hbox{\hbox
to0pt{\kern0.4\wd0\vrule height0.9\ht0\hss}\box0}}
{\setbox0=\hbox{$\scriptscriptstyle\rm C$}\hbox{\hbox
to0pt{\kern0.4\wd0\vrule height0.9\ht0\hss}\box0}}}}

\font\fivesans=cmss12 at 5pt
\font\sevensans=cmss12 at 7pt
\font\tensans=cmss12
\newfam\sansfam
\textfont\sansfam=\tensans\scriptfont\sansfam=\sevensans
\scriptscriptfont\sansfam=\fivesans
\def\sans{\fam\sansfam\tensans}
\def\bbbz{{\mathchoice {\hbox{$\sans\textstyle Z\kern-0.4em Z$}}
{\hbox{$\sans\textstyle Z\kern-0.4em Z$}}
{\hbox{$\sans\scriptstyle Z\kern-0.3em Z$}}
{\hbox{$\sans\scriptscriptstyle Z\kern-0.2em Z$}}}}

\font \bigbf=cmbx10 scaled \magstep2

\def\slash#1{#1\kern-0.65em /}
\def\dirac{{\raise0.09em\hbox{/}}\kern-0.58em\partial}
\def\Dirac{{\raise0.09em\hbox{/}}\kern-0.69em D}





\magnification=\magstep1
\parskip 4pt plus 1pt
\hoffset=0.1 truecm
\voffset=-0.25 truecm
\hsize=15.5 truecm
\vsize=24.5 truecm
\vglue 1.5cm

\centerline {\bigbf Linear Connections on Extended Space-Time}
\vskip 1.5cm

\centerline {\bf A. Kehagias}
\medskip
\centerline {\it Physics Department, National Technical University}
\centerline {\it GR-15773 Zografou, ATHENS}
\vskip 1cm

\centerline {\bf J. Madore}
\medskip
\centerline {\it Laboratoire de Physique Th\'eorique et Hautes
Energies\footnote{*}{\it Laboratoire associ\'e au CNRS.}}
\centerline {\it Universit\'e de Paris-Sud, B\^at. 211,  \ F-91405 ORSAY}
\vskip 1cm

\centerline {\bf J. Mourad}
\medskip
\centerline {\it  Laboratoire de Mod\`eles de Physique Math\'ematique}
\centerline {\it Parc de Grandmont, Universit\'e de Tours, \ F-37200 TOURS}
\vskip 1cm

\centerline {\bf G. Zoupanos}
\medskip
\centerline {\it Physics Department, National Technical University}
\centerline {\it GR-15773 Zografou, ATHENS}
\vskip 2cm

\noindent
{\bf Abstract:} \ A modification of Kaluza-Klein theory is
proposed which is general enough to admit an arbitrary finite
noncommutative internal geometry. It is shown that the existence of a
non-trival extension to the total geometry of a linear connection on
space-time places severe restrictions on the structure of the
noncommutative factor. A counter-example is given.

\vfill
\noindent
LMPM Tours 95-01
\medskip
\noindent
January, 1995
\bigskip
hep-th/9502017
\eject

\beginsection 1 Introduction

Immediately after the introduction of a noncommutative extension of
space-time to unify Higgs scalars with Yang-Mills gauge fields
(Dubois-Violette {\it et al.} 1989a) it was recognized that the
resulting geometry could be interpreted as a modification of
Kaluza-Klein theory (Dubois-Violette {\it et al.} 1989b). For later
developments we refer for example to Madore (1990), Chamseddine {\it et
al.} (1993) and Madore \& Mourad (1993). Our purpose here is to present
a formalism sufficiently general to describe any finite noncommutative
extension of space-time and to discuss the type of restrictions which it
is necessary to place on the internal differential calculus for there to
exist non-trivial extensions to the total geometry of a linear
connection on space-time.

After a preliminary description of the commutative case to fix the
notation, we give in Section~2 the general definition of what we mean by
a noncommutative extension of space-time. In Section~3 we show that this
is essentially a reformulation of a previous definition (Madore 1990,
Madore \& Mourad 1993) in a more general context. As a different
explicit example of a noncommutative geometry we recall briefly in
Section~4 the linear connection which has been recently added (Madore
{\it et al.} 1994) to the differential calculus which Connes \& Lott
(1990, 1991) have proposed to describe the Higgs sector of the Standard
Model. We use this calculus to give in Section~5 an example of a
Kaluza-Klein extension which is necessarily trivial, without Yang-Mills
potentials or scalar fields.

We refer to Bailin \& Love (1987) for an introduction to standard
Kaluza-Klein theory and, for example, to Madore \& Mourad (1993) for a
motivation of the generalization to noncommutative geometry. The
original Kaluza-Klein construction involves geometric structures on a
group $G$, on a space-time manifold $V$ and on a principal bundle $P$
over $V$. We shall distinguish the structure of the bundle by a tilde
and, when necessary, that of the manifold by a subscript $V$.

It is to be stressed that we are here concerned with a generalization of
Kaluza-Klein theory to noncommutative geometry and not with the
definition of a noncommutative version of a principal bundle. Also, as
was pointed out in the previous publications, our definition of a linear
connection makes essential use of the bimodule structure of the space of
1-forms. This accounts for the difference of our conclusions from those
of authors (Chamseddine {\it et al.} 1993, Sitarz 1994, Klim\v c\' ik
{\it et al.} 1994, Landi {\it et al.} 1994,) who define a linear
connection using the classical (Koszul 1960) formula for a covariant
derivative on an arbitrary left (or right) module. A more detailed
comparison of the two approaches is given in Sitarz (1995). We have
formulated our results where necessary directly in terms of covariant
derivatives on the bimodule structure.  In noncommutative geometry
connection forms cannot be defined in general.

Let $V$ be a differential manifold and let $(\Omega^*(V), d)$ be the
ordinary differential calculus on $V$.  A linear connection on $V$ can
be defined as a connection on the cotangent bundle to $V$. It can be
characterized (Koszul 1960) as a linear map
$$
\Omega^1(V) \buildrel D \over \longrightarrow
\Omega^1(V) \otimes_{{\cal C}(V)} \Omega^1(V)                     \eqno(1.1)
$$
which satisfies the left Leibniz rule
$$
D (f \xi) =  df \otimes \xi + f D\xi                              \eqno(1.2)
$$
for arbitrary $f \in {\cal C}(V)$ and $\xi \in \Omega^1(V)$.  Let
$\theta^\alpha$ to be a local moving frame on $V$. The connection form
$\omega^\alpha{}_\beta$ is defined in terms of the covariant derivative
of $\theta^\alpha$:
$$
D\theta^\alpha = -\omega^\alpha{}_\beta \otimes \theta^\beta.     \eqno(1.3)
$$
Because of (1.2) the covariant derivative $D\xi$ of an arbitrary element
$\xi = \xi_\alpha \theta^\alpha \in \Omega^1(V)$ can be written as
$D\xi = (D\xi_\alpha) \otimes \theta^\alpha$ where
$$
D\xi_\alpha = d\xi_\alpha - \omega^\beta{}_\alpha \xi_\beta.      \eqno(1.4)
$$
Let $\pi$ be the product in the algebra of forms. Using it one can
define the torsion form $\Theta^\alpha = (d - \pi D)\theta^\alpha$. We
shall assume that the torsion vanishes:
$$
\pi  D  = d.                                                      \eqno(1.5)
$$

Let $\sigma$ be the bilinear map of
$\Omega^1(V) \otimes_{{\cal C}(V)} \Omega^1(V)$ into itself defined by
the permutation of the two factors:
$$
\sigma (\theta^\alpha \otimes \theta^\beta)
= \theta^\beta \otimes \theta^\alpha.                             \eqno(1.6)
$$
Using $\sigma$ we can extend $D$ to the tensor algebra:
$$
D(\theta^\alpha \otimes \theta^\beta) =
D\theta^\alpha\otimes \theta^\beta +
\sigma_{12} (\theta^\alpha \otimes D\theta^\beta), \qquad
\sigma_{12} = \sigma \otimes 1.
$$
The metric can be defined as a bilinear map
$$
\Omega^1(V) \otimes_{{\cal C}(V)} \Omega^1(V)
\buildrel g \over \longrightarrow \Omega^0(V).                     \eqno(1.7)
$$
It is symmetric if
$$
g \sigma = g.                                                      \eqno(1.8)
$$
We shall assume that the connection is metric compatible:
$$
(1 \otimes g) D (\xi \otimes \eta) = dg (\xi \otimes \eta).        \eqno(1.9)
$$
Using $\pi$ one defines the curvature 2-form $\Omega^\alpha{}_\beta$:
$$
\pi_{12} D^2 \theta^\alpha
= - \Omega^\alpha{}_\beta \otimes \theta^\beta, \qquad
\pi_{12} = \pi \otimes 1.                                         \eqno(1.10)
$$
The left-linearity of the curvature,
$$
\pi_{12} D^2 (f \theta^\alpha) = f \pi_{12} D^2 \theta^\alpha,    \eqno(1.11)
$$
is a consequence of the identity
$$
\pi (\sigma + 1) = 0.                                             \eqno(1.12)
$$

The module $\Omega^1(V)$ has a natural structure as a right
${\cal C}(V)$-module and the right Leibniz rule is determined using the
fact that ${\cal C}(V)$ is a commutative algebra:
$$
D (\xi f) =  D (f \xi).                                           \eqno(1.13)
$$
Using $\sigma$ this can also be written in a form
$$
D(\xi f) = \sigma (\xi \otimes df) + (D\xi) f.                    \eqno(1.14)
$$
which can be used in noncommutative geometries.

Let now $(\Omega^*, d)$ designate a general differential calculus. It was
suggested by Dubois-Violette \& Michor (1994) and by Mourad (1994) that
an essential ingredient in the definition of a linear connection over
$\Omega^*$ is a generalized symmetry operation $\sigma$:
$$
\Omega^1 \otimes_{\Omega^0}  \Omega^1
\buildrel \sigma \over \longrightarrow
\Omega^1 \otimes_ {\Omega^0} \Omega^1.                            \eqno(1.15)
$$
It can be shown that $\sigma$ is bilinear and in some examples which
have been considered it can also be shown that $\sigma$ is essentially
unique (Dubois-Violette {\it et al.} 1994, Madore {\it et al.} 1994).
In general $\sigma^2 \neq 1$ but if one supposes that $\sigma$ satisfies
the Hecke relation then one can define the exterior algebra and the
symmetric algebra as subalgebras of the tensor algebra.

A covariant derivative is defined as a linear map
$$
\Omega^1 \buildrel D \over \longrightarrow
\Omega^1 \otimes_{\Omega^0} \Omega^1                              \eqno(1.16)
$$
which satisfies the Leibniz rules (1.2) and (1.14) but with $\sigma$
given by (1.15). The condition that the linear connection be
torsion-free is given as before by (1.5).  A metric $g$ is a
$\Omega^0$-bilinear map
$$
\Omega^1 \otimes_ {\Omega^0} \Omega^1
\buildrel g \over \longrightarrow \Omega^0.                        \eqno(1.17)
$$
It is symmetric if (1.8) is satisfied. The covariant derivative is
metric compatible if (1.9) is satisfied for $\xi$, $\eta$ elements of
$\Omega^1$.

Consider the commutative Kaluza-Klein construction.  Suppose that $P$ is
a general $G$-bundle over $V$ and let $p$ be the projection of $P$ onto
$V$, Define $\tilde \theta^\alpha = p^*(\theta^\alpha)$.  Suppose that
there are Yang-Mills fields present and let $\tilde \theta^a$ be the
components of a Yang-Mills connection with curvature $F$.  We can define
a metric on $P$ by requiring that
$\tilde \theta^i = (\tilde \theta^\alpha, \tilde \theta^a)$ be an
orthonormal moving frame. A covariant derivative on $P$ is given by
$$
\eqalign{
&\tilde D\tilde \theta^\alpha = - \omega^\alpha{}_\beta \otimes
\tilde \theta^\beta + \Gamma^\alpha,                \cr
&\tilde D\tilde \theta^a
= -{1\over 2}C^a{}_{bc} \tilde \theta^b \otimes \tilde \theta^c + F^a
}                                                                 \eqno(1.18)
$$
where $\Gamma^\alpha \in \Omega^1(P) \otimes \Omega^1(P)$ is determined
by the condition
$$
\tilde \pi \Gamma^\alpha = 0                                      \eqno(1.19)
$$
that the connection be torsion-free and the condition
$$
\Gamma^\alpha \otimes \tilde \theta^a
+ \tilde \sigma_{12} (\tilde \theta^\alpha \otimes F^a) = 0       \eqno(1.20)
$$
that it be metric compatible. In the above formulae, $\tilde \pi$ is the
product in $\tilde \Omega^*$ and $\tilde \sigma$ is the natural
extension of (1.6) to $\tilde \Omega^1 \otimes \tilde \Omega^1$.

Our purpose is to reformulate the Kaluza-Klein construction in
sufficient generality that the group manifold can be replaced by an
arbitrary noncommutative geometry.  For this we must be able to replace
the $\tilde \theta^a$ by 1-forms $\xi$ in some noncommutative
differential calculus.  A Yang-Mills potential will be a 1-form on
space-time with values in the algebra which describes the noncommutative
geometry. In a typical application the anti-hermitian elements of this
algebra can be identified with the Lie algebra of a subgroup of a
general linear group.

\beginsection 2 The general theory

In the most general context a Kaluza-Klein theory can be based on
an extension $(\tilde \Omega^*, \tilde d)$ of $(\Omega^*(V), d)$ defined
by a differential algebra $\tilde \Omega^*$ with an imbedding
$$
\Omega^*(V)
\buildrel i \over \longrightarrow \tilde \Omega^*                  \eqno(2.1)
$$
and a differential $\tilde d$ which coincides with $d$ on $\Omega^*(V)$.
In the previous section $i$ was given by $i = p^*$.  In particular we
shall define $\tilde \theta^\alpha = i (\theta^\alpha)$.  We shall here
only consider the special case with
$$
\tilde \Omega^* = \Omega^*(V) \otimes_\bbbc \Omega^*,              \eqno(2.2)
$$
where $(\Omega^*, d)$ is a finite-dimensional differential calculus over
the complex numbers. The algebra $\tilde \Omega^0$ could be, for
example, the algebra of automorphisms of a trivial vector bundle over
$V$. The algebra $\Omega^0(V)$ is imbedded in the center $\tilde Z^0$ of
$\tilde \Omega^0$. In Section~5 we shall consider a case where
$\tilde Z^0$ can be identified with the algebra of functions on two
copies of $V$.

Consider first the case of a general (algebraic) tensor product
$$
\tilde \Omega^* = \Omega^{\prime *} \otimes_\bbbc \Omega^{\prime\prime *}
                                                                   \eqno(2.3)
$$
of two arbitrary differential calculi $\Omega^{\prime *}$ and
$\Omega^{\prime\prime *}$ with corresponding covariant derivatives
$D^\prime$ and $D^{\prime\prime}$. We have then two bilinear maps
$$
\Omega^{\prime 1} \buildrel D^\prime \over \longrightarrow
\Omega^{\prime 1} \otimes_{\Omega^{\prime 0}} \Omega^{\prime 1},    \qquad
\Omega^{\prime\prime 1} \buildrel D^{\prime\prime} \over \longrightarrow
\Omega^{\prime\prime 1} \otimes_{\Omega^{\prime\prime 0}}
\Omega^{\prime\prime 1},                                           \eqno(2.4)
$$
from which we wish to construct an extension
$$
\tilde \Omega^1 \buildrel \tilde D \over \longrightarrow
\tilde \Omega^1 \otimes_{\tilde \Omega^0} \tilde \Omega^1.         \eqno(2.5)
$$
This is the most general possible formulation of the bosonic part of the
Kaluza-Klein construction under the condition (2.3).

The 1-forms $\tilde \Omega^1$ can be written as a direct sum
$$
\tilde \Omega^1 = \tilde \Omega^1_h \oplus \tilde \Omega^1_c       \eqno(2.6)
$$
where, in the traditional language of Kaluza-Klein theory, $\Omega^1_h$
is the horizontal component of the 1-forms. It can be defined as the
$\tilde \Omega^0$-module generated by the image of $\Omega^{\prime 1}$
in $\tilde \Omega^1$ under the generalization of the map (2.1); it is
given by
$$
\tilde \Omega^1_h =
\Omega^{\prime 1} \otimes_\bbbc \Omega^{\prime\prime 0}.            \eqno(2.7)
$$
The $\tilde \Omega^1_c$ is a complement of $\tilde \Omega^1_h$ in
$\tilde \Omega^1$. Because of the Ansatz (2.3) such a complement always
exists. We shall choose
$$
\tilde \Omega^1_c =
\Omega^{\prime 0} \otimes_\bbbc \Omega^{\prime\prime 1}.            \eqno(2.8)
$$
The horizontal component of the 1-forms can be expected to have
a more general significance whereas the existence of the complement
depends on the Ansatz (2.3).

Let $f^\prime \in \Omega^{\prime 0}$
($f^{\prime\prime} \in \Omega^{\prime\prime 0}$) and
$\xi^{\prime} \in \Omega^{\prime 1}$
($\xi^{\prime\prime} \in \Omega^{\prime\prime 1})$.  Then it follows
from the definition of the product in the tensor product that
$$
f^\prime \xi^{\prime\prime} = \xi^{\prime\prime} f^\prime, \qquad
f^{\prime\prime} \xi^\prime = \xi^\prime f^{\prime\prime}.
$$
Hence from (1.12) one concludes that the extension $\tilde\sigma$ of
$\sigma^\prime$ and $\sigma^{\prime\prime}$ which is part of the
definition of $\tilde D$ is given by
$$
\tilde\sigma(\xi^\prime \otimes \eta^{\prime\prime}) =
\eta^{\prime\prime} \otimes \xi^\prime,                     \qquad
\tilde\sigma(\xi^{\prime\prime} \otimes \eta^\prime) =
\eta^\prime \otimes \xi^{\prime\prime}.                            \eqno(2.9)
$$
{}From these one deduces the constraints
$$
f^\prime \tilde D \xi^{\prime\prime} =
(\tilde D \xi^{\prime\prime}) f^\prime,\qquad
f^{\prime\prime} \tilde D \xi^\prime =
(\tilde D \xi^\prime) f^{\prime\prime}                            \eqno(2.10)
$$
on $\tilde D$. These are trivially satisfied if
$$
\tilde D\xi^\prime = D^\prime \xi^\prime,                   \qquad
\tilde D\xi^{\prime\prime} = D^{\prime\prime} \xi^{\prime\prime}. \eqno(2.11)
$$

Using the decomposition (2.6) one sees that the covariant derivative
(2.5) takes its values in the sum of 4 spaces, which can be written in
the form
$$
\eqalign{
&\tilde \Omega^1_h \otimes_{\tilde \Omega^0} \tilde \Omega^1_h =
(\Omega^{\prime 1} \otimes_{\Omega^{\prime 0}} \Omega^{\prime 1})
\otimes_\bbbc \Omega^{\prime\prime 0},                             \cr
&\tilde \Omega^1_h \otimes_{\tilde \Omega^0} \tilde \Omega^1_v =
\Omega^{\prime 1} \otimes_\bbbc \Omega^{\prime\prime 1},           \cr
&\tilde \Omega^1_v \otimes_{\tilde \Omega^0} \tilde \Omega^1_h =
\Omega^{\prime\prime 1}  \otimes_\bbbc \Omega^{\prime 1},          \cr
&\tilde \Omega^1_v \otimes_{\tilde \Omega^0} \tilde \Omega^1_v =
\Omega^{\prime 0} \otimes_\bbbc (\Omega^{\prime\prime 1}
\otimes_{\Omega^{\prime\prime 0}} \Omega^{\prime\prime 1}).        \cr
}                                                                  \eqno(2.12)
$$
Let $Z^{\prime 0}$ ($Z^{\prime\prime 0}$) be the center of
$\Omega^{\prime 0}$ ($\Omega^{\prime\prime 0}$) and let  $Z^{\prime 1}$
($Z^{\prime\prime 1}$) be the vector space of elements of
$\Omega^{\prime 1}$ ($\Omega^{\prime\prime 1}$) which commute with
$\Omega^{\prime 0}$ ($\Omega^{\prime\prime 0}$). Then $Z^{\prime 1}$
($Z^{\prime\prime 1}$) is a bimodule over $Z^{\prime 0}$
($Z^{\prime\prime 0}$). Let $Z^{\prime 2}$ ($Z^{\prime\prime 2}$)
be the elements of
$\Omega^{\prime 1} \otimes_{\Omega^{\prime 0}} \Omega^{\prime 1}$
($\Omega^{\prime\prime 1} \otimes_{\Omega^{\prime\prime 0}}
\Omega^{\prime\prime 1}$) which commute with $\Omega^{\prime 0}$
($\Omega^{\prime\prime 0}$). Then
$$
Z^{\prime 1} \otimes_{Z^{\prime 0}} Z^{\prime 1}
\subset Z^{\prime 2},                                    \qquad
Z^{\prime\prime 1} \otimes_{Z^{\prime\prime 0}} Z^{\prime\prime 1}
\subset Z^{\prime\prime 2},
$$
but in general the two sides are not equal.  From (2.10) we see then
that
$$
\eqalign{
&\tilde D\xi^\prime \in
(\Omega^{\prime 1} \otimes_{\Omega^{\prime 0}} \Omega^{\prime 1})
\otimes_\bbbc Z^{\prime\prime 0}                          \oplus
\Omega^{\prime 1} \otimes_\bbbc Z^{\prime\prime 1}        \oplus
Z^{\prime\prime 1} \otimes_\bbbc \Omega^{\prime 1}        \oplus
\Omega^{\prime 0} \otimes_\bbbc Z^{\prime\prime 2},       \cr
&\tilde D\xi^{\prime\prime} \in
Z^{\prime 0} \otimes_\bbbc (\Omega^{\prime\prime 1}
\otimes_{\Omega^{\prime\prime 0}} \Omega^{\prime\prime 1})\oplus
\Omega^{\prime\prime 1} \otimes_\bbbc Z^{\prime 1}        \oplus
Z^{\prime 1} \otimes_\bbbc \Omega^{\prime\prime 1}        \oplus
Z^{\prime 2} \otimes_\bbbc \Omega^{\prime\prime 0}.
}                                                                  \eqno(2.13)
$$
In the relevant special case with
$\Omega^{\prime *} = \Omega^*(V)$, we shall have
$$
Z^{\prime 0} = \Omega^{\prime 0},  \qquad
Z^{\prime 1} = \Omega^{\prime 1},                                  \eqno(2.14)
$$
and so (2.13) places no restriction on $\tilde D\xi^{\prime\prime}$.
However if
$$
Z^{\prime\prime 1} = 0, \qquad
Z^{\prime\prime 2} = 0,                                            \eqno(2.15)
$$
one finds the constraint
$$
\tilde D\xi^\prime = D^\prime \xi^\prime.                          \eqno(2.16)
$$

We shall impose the condition that the connections be metric and without
torsion although these might be considered rather artificial conditions
on the vertical component of the 1-forms. We have then two bilinear maps
$$
\Omega^{\prime 1} \otimes_{\Omega^{\prime 0}} \Omega^{\prime 1}
\buildrel g^\prime \over \longrightarrow \Omega^{\prime 0}, \qquad
\Omega^{\prime\prime 1} \otimes_{\Omega^{\prime\prime 0}}
\Omega^{\prime\prime 1}
\buildrel g^{\prime\prime} \over \longrightarrow
\Omega^{\prime\prime 0},                                           \eqno(2.17)
$$
which satisfy the compatibility condition (1.9), from which we must
construct an extension
$$
\tilde \Omega^1 \otimes_{\tilde \Omega^0} \tilde \Omega^1
\buildrel \tilde g \over \longrightarrow \tilde \Omega^0
$$
which satisfies also (1.9). From the decomposition (2.12) one sees that
$\tilde g$ will be determined by two bilinear maps
$$
\Omega^{\prime 1} \otimes_\bbbc \Omega^{\prime\prime 1}
\buildrel g_1 \over \longrightarrow \tilde \Omega^0,  \qquad
\Omega^{\prime\prime 1} \otimes_\bbbc \Omega^{\prime 1}
\buildrel g_2 \over \longrightarrow \tilde \Omega^0.
                                                                   \eqno(2.18)
$$
If $\tilde g$ is symmetric then from (2.9) it follows that
$$
g_2 = g_1 \tilde \sigma.
$$
In general it is to be expected that if the connection is metric and
without torsion, the conditions (2.15) will place constraints also on the
covariant derivative $\tilde D\xi^{\prime\prime}$.

In the relevant special case with $\Omega^{\prime *} = \Omega^*(V)$ one
can define a metric $i^* \tilde g$ on $V$ by
$$
i^* \tilde g(\theta^\alpha, \theta^\beta)
  = \tilde g(\tilde\theta^\alpha, \tilde\theta^\beta),     \qquad
    \tilde\theta^\alpha = i(\theta^\alpha).
$$
To maintain contact with the commutative construction of the previous
section we suppose in this case that
$$
i^* \tilde g = g_V                                                 \eqno(2.19)
$$
where $g_V$ is a metric on $V$.

A classical fermionic field associated to a differential calculus
$(\Omega^*, d)$ lies in a left $\Omega^0$-module ${\cal H}$ and its
dynamics are governed by a Dirac operator $\Dirac$ which is a
left-linear map of ${\cal H}$ into itself. In Kaluza-Klein theory we
have then given a $\Dirac^\prime$ on a left $\Omega^{\prime 0}$-module
${\cal H}^\prime$ and a $\Dirac^{\prime\prime}$ on a left
$\Omega^{\prime\prime 0}$-module ${\cal H}^{\prime\prime}$ from which we
must construct a Dirac operator $\tilde \Dirac$ on the left
$\tilde \Omega^0$-module
$$
\tilde {\cal H} = {\cal H}^\prime \otimes_\bbbc {\cal H}^{\prime\prime}.
                                                                   \eqno(2.20)
$$

\beginsection 3 An example             

In this section we shall mention an example with a noncommutative
internal structure which leads to a non-trivial Kaluza-Klein extension.
It is based (Madore \& Mourad 1993) on the algebra of $n\times n$
matrices with a differential calculus derived from derivations.
A basis $e_a$ of the derivations of $M_n$ is provided by
$n^{2}-1$ independent traceless anti-selfadjoint matrices, $\lambda_{a}$:
$$
e_{a}(f)=ad_{e_{a}}f=[\lambda_{a},f],\quad f\in M_{n}.             \eqno(3.1)
$$
The set of 1-forms, $\Omega^{1}$, is a $M_{n}$-bimodule freely generated
by the duals $\theta^{a}$ of $e_{a}$:
$$
\theta^{a}(e_{b})=\delta^{a}_{b}.                                  \eqno(3.2)
$$
We shall need the exterior derivative of these 1-forms:
$$
d\theta^{a}=-{1\over 2}C^{a}{}_{bc}\theta^{b}\theta^{c},           \eqno(3.3)
$$
where $C^{a}{}_{bc}$ are the $SU_{n}$ structure constants with respect
to the basis $\lambda_{a}$. We shall also need the important
property
$$
f\theta^{a}=\theta^{a}f, \quad \forall f\in M_{n}.                  \eqno(3.4)
$$
The generalized symmetry operation $\sigma$
was given by Madore {\it et al.} (1994):
$$
\sigma(\theta^{a}\otimes\theta^{b})=\theta^{b}\otimes\theta^{a}.    \eqno(3.5)
$$

A general element, $\alpha$ of $\tilde \Omega^{1}$ can be written as
a sum  $\alpha= A+\xi$ where
$A\in \tilde \Omega^{1}_{h}= \Omega^{1}(V)\otimes_{\bbbc}M_{n}$ is a 1-form
on $V$ with values in $M_{n}$ and
$\xi\in \tilde \Omega^{1}_{c} = \Omega^{0}(V)\otimes_{\bbbc}\Omega^{1}(M_{n})$.
Introduce $\theta^{i}=(\theta^{\alpha},\theta^{a})$. The generalized
symmetry operation $\tilde\sigma$ is given by
$$
\tilde \sigma( \theta^i \otimes  \theta^j) = \theta^j \otimes \theta^i.
                                                                    \eqno(3.6)
$$
{}From the property (3.4)we find that
$$
Z''^{0}=\{1\},    \qquad        Z''^{1}=\{\theta^{a}\},   \qquad
Z''^{2}=\{\theta^{a}\otimes\theta^{b}\}.                            \eqno(3.7)
$$
We can write then the most general covariant derivative
as
$$
\tilde D \theta^{i} = -\Gamma^{i}{}_{jk} \theta^{j} \otimes \theta^{k},
                                                                    \eqno(3.8)
$$
where
$$
\Gamma^{i}{}_{jk} \in \Omega^{0}(V).                                \eqno(3.9)
$$

{}From the condition (1.5) one finds
$$
d \theta^{i} = -\Gamma^{i}{}_{jk} \theta^{j} \wedge  \theta^{k}.   \eqno(3.10)
$$
Therefore
$$
\matrix{
\Gamma^{a}{}_{bc}-\Gamma^{a}{}_{cb}=C^{a}{}_{bc},                  \hfill
&\Gamma^{a}{}_{b\alpha}=\Gamma^{a}{}_{\alpha b},                   \hfill\cr
\Gamma^{\alpha}{}_{a\beta}=\Gamma^{\alpha}{}_{\beta a},            \hfill
&\Gamma^{\alpha}{}_{a b}=\Gamma^{\alpha}{}_{b a},                  \hfill\cr
\Gamma^{\alpha}{}_{\beta\gamma}-\Gamma^{\alpha}{}_{\gamma\beta} =
\omega^{\alpha}{}_{\beta\gamma}-\omega^{\alpha}{}_{\gamma\beta}.   \hfill
}
                                                                   \eqno(3.11)
$$
where
$$
\omega^\alpha{}_\beta = \omega^\alpha{}_{\gamma\beta}\theta^\gamma
$$
is a torsion-free connection 1-form on $V$.

The metric is given by $g^{ij}=g(\theta^{i} \otimes \theta^{j})$.  Since
the $\theta^{i} \otimes \theta^{j}$ commute with all elements of the
algebra, the $g^{ij}$ are complex functions.  The reality condition
$$
(g(\xi\otimes\eta))^* = g(\sigma(\eta^*\otimes\xi^*))              \eqno(3.12)
$$
imposes that the $g^{ij}$ be real and the symmetry condition
$g = g\circ\sigma$ yields $g^{ij}=g^{ji}$. A non-degenerate metric is
characterized by an invertible $g^{ij}$.

The compatibility condition (1.9) becomes
$$
dg^{ij} = -\Gamma^{i}{}_{kl} \theta^{k}g^{lj}
          -\Gamma^{j}{}_{kl} \theta^{k}g^{il}.                     \eqno(3.13)
$$
This condition together with the vanishing-torsion condition determines
the linear connection. If we define
$$
T^i{}_{jk} = {1 \over 2}(\Gamma^{i}{}_{jk}-\Gamma^{i}{}_{kj})
$$
we find
$$
\Gamma^{i}{}_{jk}=
{1 \over 2}g^{il}(e_{k}g_{jl}+e_{j}g_{kl}-e_{l}g_{jk})
+ T^i{}_{jk} - T_{kj}{}^i - T_{jk}{}^i,                            \eqno(3.14)
$$
where $g_{ij}$ in the inverse of $g^{ij}$.  Note that since the $g_{ij}$
are functions we have
$$
e_{a}g_{ij}=0.                                                     \eqno(3.15)
$$
Using the metric we can construct from the basis $\theta^i$ an
orthonormal basis $\tilde \theta^i$ such as was used in the
topologically more complicated situation considered in Section~1.

The difference between the present calculations and the previous (Madore
\& Mourad 1993) is that we have here considered the most general
connection and metric on the tensor product algebra and have shown that
our postulates on the covariant derivative and the metric lead
necessarily to components in the basis $\theta^{i}$ which are functions
of $V$ alone. Also, in the previous calculations the metric was
considered an element of $\Omega^{1} \otimes \Omega^{1}$. Had we used
this definition here the fact that the $g_{ij}$ are complex function
would have arisen as a consequence of the compatibility condition
$\tilde D g=0$.

\beginsection 4 The Connes-Lott geometry

As an example we consider the differential calculus which has been
proposed by Connes \& Lott (1990, 1991) to describe the Higgs sector of
the Standard Model. As mentioned in the previous section, to define a
linear connection on the bimodule $\Omega^1$, we must suppose the
existence of a bilinear map (1.15) to replace the usual symmetry
operation which is used to define differential forms. In the example we
consider here $\sigma^2 = 1$ if and only if the unique connection is
metric compatible.

The Connes-Lott geometry is based on a differential calculus over an
algebra of matrices with a differential defined by a graded commutator
(Connes 1986).  Consider the matrix algebra $M_n$ with a $\bbbz_2$
grading. One can define on $M_n$ a graded derivation $\hat d$ by the
formula
$$
\hat d f = - [ \theta , f],                                        \eqno(4.1)
$$
where $\theta$ is an arbitrary anti-hermitian odd element
and the commutator is taken as a graded commutator. We find that
$\hat d\theta = -2\theta^2$ and for any $\alpha \in M_n$,
$$
\hat d^2 \alpha = [ \theta^2, \alpha ].                            \eqno(4.2)
$$
The $\bbbz_2$ grading of $M_n$ can be expressed as the direct sum
$M_n = M_n^+ \oplus M_n^-$ where $M_n^+$
($M_n^-$) are the even (odd) elements of $M_n$.  It can be
induced from a decomposition $\bbbc^n = \bbbc^l \oplus \bbbc^{n-l}$ for
some integer $l$.  The elements of $M_n^+$ are diagonal with
respect to the decomposition; the elements of $M_n^-$ are
off-diagonal.

It is possible to construct over $M_n^+$ a differential algebra
$\Omega^* = \Omega^*(M_n^+)$ (Connes \& Lott 1991).
Let $\Omega^0 = M^+_n$ and let
$\Omega^1 \equiv \overline{d\Omega^0} \subset M^-_n$ be the
$M^+_n$-bimodule generated by the image of $\Omega^0$ in $M^-_n$
under $\hat d$.  Define
$$
\Omega^0 \buildrel d \over \longrightarrow \Omega^1              \eqno(4.3)
$$
using directly (4.1): $d = \hat d$. Let
$\overline{d\Omega^1}$ be the $M^+_n$-module generated by the
image of $\Omega^1$ in $M^+_n$ under $\hat d$.  It would be natural
to try to set $\Omega^2 = \overline{d\Omega^1}$ and define
$$
\Omega^1 \buildrel d \over \longrightarrow \Omega^2              \eqno(4.4)
$$
using once again (4.1). Every element of $\Omega^1$ can be
written as a sum of elements of the form $f_0 \hat d f_1$. If we attempt
to define an application (4.6) using again directly (4.3),
$$
d (f_0 \hat d f_1) = \hat df_0 \hat df_1 + f_0 \hat d^2 f_1,      \eqno(4.5)
$$
then we see that in general $d^2$ does not vanish. To remedy this
problem we eliminate simply the unwanted terms. Let  ${\rm Im}\,\hat d^2$
be the submodule of $\overline{d\Omega^1}$ consisting of those elements
which contain a factor which is the image of $\hat d^2$ and define
$\Omega^2$ by
$$
\Omega^2 = \overline{d\Omega^1} / {\rm Im}\,\hat d^2.             \eqno(4.6)
$$
Then by construction the second term on the right-hand side of (4.5)
vanishes as an element of $\Omega^2$ and we have a well defined map
(4.4) with $d^2 = 0$.  This procedure can be continued to arbitrary
order by iteration.  For each $p \geq 2$ we let ${\rm Im}\,\hat d^2$ be
the submodule of $\overline{d\Omega^{p-1}}$ defined as above and
we define $\Omega^p$ by
$$
\Omega^p = \overline{d\Omega^{p-1}} / {\rm Im}\,\hat d^2.
                                                                  \eqno(4.7)
$$
Since $\Omega^p \Omega^q \subset \Omega^{p+q}$ the complex
$\Omega^*$ is a differential algebra.  The $\Omega^p$ need not
vanish for large values of $p$. In fact if $\theta^2 \propto 1$ we see
that $\hat d^2 = 0$ and the sequence defined by (4.9) never
stops.  However $\Omega^p \subseteq M^+_n (M^-_n)$ for
$p$ even (odd) and so it stabilizes for large $p$.

We shall consider in some detail the case $n=3$ with the grading
defined by the decomposition $\bbbc^3 = \bbbc^2 \oplus \bbbc$.
The most general possible form for $\theta$ is
$$
\theta = \eta_1 - \eta_1^*                                      \eqno(4.8)
$$
where
$$
\eta_1 = \pmatrix{0  &  0  &     a  \cr
                  0  &  0  &     b  \cr
                  0  &  0  &     0}.                            \eqno(4.9)
$$
Without loss of generality we can choose the euclidean 2-vector
$\eta_{1i}$ of unit length. The general construction yields
$\Omega^0 = M_3^+ = M_2 \times M_1$ and $\Omega^1 = M_3^-$
but after that the quotient by elements of the form ${\rm Im}\,\hat d^2$
reduces the dimension. One finds $\Omega^2 = M_1$ and
$\Omega^p = 0$ for $p\geq 3$. Let $e$ be the unit of $M_1$. It generates
$\Omega^2$ and can also be considered as an element of $\Omega^0$.

To form a basis for $\Omega^1$ we must introduce a second matrix
$\eta_2$.  It is convenient to choose it of the same form as $\eta_1$.
We have then in $\Omega^2$ the identity
$$
\eta_i \eta^*_j = 0.
$$
We can uniquely fix $\eta_2$ by requiring that
$$
\eta^*_i \eta_j = \delta_{ij} e.                                \eqno(4.10)
$$
It follows that
$$
d\eta_1 = e,   \qquad  d\eta_2 = 0.
$$
There is a (unique) unitary element of $M_2 \subset M_3^+$ which
exchanges $\eta_1$ and $\eta_2$:
$$
\eta_2 = u \eta_1, \qquad \eta_1 = - u \eta_2.                  \eqno(4.11)
$$
We have also
$$
\eta_2 u = 0, \qquad \eta_1 u = 0.                              \eqno(4.12)
$$

The vector space of 1-forms is of dimension 4 over the complex numbers.
The dimension of $\Omega^1 \otimes_\bbbc\Omega^1$ is equal to 16 but the
dimension of the tensor product $\Omega^1 \otimes_{M_3^+}\Omega^1$ is
equal to 5.  We shall choose
$$
\eta_{ij} = \eta_i \otimes \eta^*_j, \qquad
\zeta = \eta^*_1 \otimes \eta_1                                 \eqno(4.13)
$$
as independent basis elements. The most general $\sigma$ is given by
$$
\sigma(\eta_{11}) = \mu \eta_{11}, \qquad
\sigma(\zeta) = - \zeta.                                         \eqno(4.14)
$$
There is an imbedding $\iota$ of $\Omega^2$ into
$\Omega^1 \otimes \Omega^1$ given by $e \mapsto \iota(e) = \zeta$. If
$\mu = 1$ we can write $\sigma = 1 - 2 \iota \pi$.

Let $\eta$ be a general element of $\Omega^1$.  The unique covariant
derivative (Madore {\it et al.} 1994) is given by
$$
D\eta = \sigma (\eta \otimes \theta) - \theta \otimes \eta.      \eqno(4.15)
$$
If
$$
\mu = 1                                                          \eqno(4.16)
$$
it is compatible with the (non-symmetric) metric
$$
g(\eta_{ij}) =  \eta_i \eta^*_j \qquad g(\zeta) = - e,           \eqno(4.17)
$$
where the right-hand sides are considered as elements of $M^+_3$.  We
have put the single overall scale factor equal to one.
If $\mu = 1$ then it is possible to extend the definition of the complex
conjugation to the tensor product by the formula
$$
(\xi \otimes \eta)^* = \sigma (\eta^* \otimes \xi^*).            \eqno(4.18)
$$
With this definition the covariant derivative (4.15)) is real and the
metric (4.17) is real on $\eta_{ij}$ and imaginary on $\zeta$.

The 1-form $\theta$ is a basis for $\Omega^1$ as a bimodule and so one
can think of the geometry as being `one dimensional'. But there is
nonvanishing curvature, with one component $R_{(\theta)}$ given by
$$
R_{(\theta)} = 2.                                                \eqno(4.19)
$$

\beginsection 5 A counter-example

We can use the example of the preceding section to construct a geometry
which does not possess a non-trivial linear connection in the sense of
Kaluza and Klein.  In the calculations based on (2.3) we suppose then
that
$$\Omega^{\prime *} = \Omega^*(V), \qquad
\Omega^{\prime\prime *} = \Omega^*
$$
with $\Omega^*$ given by (4.7).  A general element
$\alpha \in \tilde \Omega^1$ can be written as a sum $\alpha = A + \xi$
where $A \in \tilde \Omega^1_h = \Omega^1(V) \otimes_\bbbc M^+_3$ is a
1-form on $V$ with values in $M^+_3$ and
$\xi \in \tilde \Omega^1_c = \Omega^0(V) \otimes_\bbbc M^-_3$ can be
considered as a set of 4 scalar fields. From (2.9) we see that the
generalized symmetry operation $\tilde\sigma$ is given by
$$
\tilde\sigma (\theta^\alpha \otimes \theta^\beta) =
\theta^\beta \otimes \theta^\alpha,                      \qquad
\tilde\sigma (\theta^\alpha \otimes \xi) =
\xi \otimes \theta^\alpha,                               \qquad
\tilde\sigma (\xi \otimes \theta^\alpha) =
\theta^\alpha \otimes \xi,                                         \eqno(5.1)
$$
with
$\tilde\sigma (\xi \otimes \eta) = \sigma (\xi \otimes \eta)$ given
by (4.14) with $\mu = 1$.

Consider the unit $e$ defined in Section~3 and set $\epsilon = 1 - 2e$.
Then $\epsilon^2 = 1$ and the elements of $M^+_3$ ($M^-_3$)
(anti-)commute with $\epsilon$.  It is easy to see that
$$
Z^{\prime\prime 0} = \{1, \epsilon \},\qquad
Z^{\prime\prime 1} = 0,    \qquad Z^{\prime\prime 2} = \{ \zeta \}.\eqno(5.2)
$$
The most general covariant derivative $\tilde D$ which satisfies the
constraints (2.13) is of the form
$$
\eqalign{
&\tilde D\theta^\alpha = - \omega^\alpha{}_\beta \otimes \theta^\beta
                + \Gamma^\alpha \zeta,             \cr
&\tilde D\theta = f \zeta + \Gamma +  F,
}                                                                 \eqno(5.3)
$$
where
$$
\matrix
{
\omega^\alpha{}_\beta \in \Omega^1(V) \otimes Z^{\prime\prime 0},  \hfill
&\Gamma^\alpha \in \Omega^0(V), \quad f \in \Omega^0(V),           \hfill\cr
\Gamma \in M^-_3 \otimes_\bbbc \Omega^1(V)                         \oplus
\Omega^1(V) \otimes_\bbbc M^-_3,                                   \hfill
&F \in \Omega^1(V) \otimes_{\Omega^0(V)} \Omega^1(V) \otimes M^+_3.
}                                                                  \eqno(5.4)
$$
The notation has been chosen here to mimic that of (1.18).

If one takes the covariant derivative of the identity
$$
\epsilon \theta + \theta \epsilon = 0                              \eqno(5.5)
$$
one finds, using (1.2) and (1.14), that
$$
\epsilon \tilde D\theta + (\tilde D\theta) \epsilon + 4\zeta = 0,  \eqno(5.6)
$$
from which it follows that
$$
f = 2, \qquad  F = 0.                                              \eqno(5.7)
$$

The identity (5.5) is satisfied by any element of $M^-_3$.  To within a
factor which lies in $Z^{\prime\prime 0}$, the 1-form $\theta$ can be
characterized by equations of the form
$$
u \theta = 0, \qquad \theta u^* = 0,
$$
where $u \in M^+_3$. Equations of this sort would be impossible in a
commutative geometry since $\theta$ generates $\Omega^1$.  If one takes
their covariant derivative one sees that $\Gamma$ must be of the form
$$
\Gamma = A \otimes \theta + \theta  \otimes B,
$$
where $A = A_1 + \epsilon A_2$ and $B = B_1 + \epsilon B_2$ are elements of
$\Omega^1(V) \otimes Z^{\prime\prime 0}$.  From (5.5) and the condition
(1.5) that the torsion vanish one concludes that
$$
B_1 = A_1,  \quad  B_2 = - A_2,  \qquad \Gamma^\alpha = 0.        \eqno(5.8)
$$
Therefore the most general torsion-free connection is given by
$$
\eqalign{
&\tilde D\theta^\alpha =
- \omega^\alpha{}_\beta \otimes \theta^\beta,         \cr
&\tilde D\theta = 2 \zeta
+ A_\alpha (\theta^\alpha \otimes \theta
+ \theta \otimes \theta^\alpha),
}                                                                  \eqno(5.9)
$$
with
$\omega^\alpha{}_\beta$ and $ A_\alpha$ elements of
$\Omega^1(V) \otimes Z^{\prime\prime 0}$.

{}From the general discussion of Section~2 the extension $\tilde g$ of the
metric is determined by two functions $g_1 (\theta^\alpha, \theta)$
and $g_2 (\theta, \theta^\alpha)$ on $V$ with values in $M^+_3$. But
from the relations (5.5) and the supposed bilinearity one concludes that
they must vanish:
$$
g_1 = 0, \qquad g_2 = 0.                                           \eqno(5.10)
$$
The metric $\tilde g$ is given then by the metric $g^\prime$ on
$V$ with values in $Z^{\prime\prime 0}$ and the metric (4.17) with a
possible function on $V$ as extra overall scale factor.

The condition (1.9) that the extended connection be compatible with the
extended metric implies that the extension of the covariant derivative
is trivial:
$$
\omega^\alpha{}_\beta \in \Omega^1(V), \qquad A = 0.               \eqno(5.11)
$$
The metric $\tilde g$ is given by a metric $g^\prime = g_V$ on $V$
and the metric (4.17) with no extra scalar field.

\beginsection 6 Conclusions

We have proposed a general noncommutative extension of Kaluza-Klein
theory and we have discussed the type of restrictions which must be
placed on the supplementary structure to render possible a non-trivial
extension of a linear connection. Most important of these is the
existence of 1-forms which commute with the algebra. In the formulation
of Kaluza-Klein theory using a differential calculus based on
derivations this condition is satisfied. We have presented an example
where it is not. In situations where the imbedding (2.1) cannot be
reduced to the product (2.2) and which would be the noncommutative
analogue of non-trivial principle bundles over $V$ then the
decomposition (2.6) is no longer possible. The conclusions of Section~5
remain however valid since they are local in $V$.

\parskip 7pt plus 1pt
\parindent=0cm
{\it Acknowledgment:}\ One of the authors (J. Ma.) would like to thank
M. Dubois-Violette and A. Sitarz for enlightening conversations. This
research was partially subsidized by the CEC Science project No.
SCI-CT91-0729.
\vskip 1cm
\vfill\eject

\beginsection References

Bailin D., Love A. 1987, {\it Kaluza-Klein theories}, Rep. Prog. Phys.
{\bf 50} 1087.

Chamseddine A.H., Felder G., Fr\"ohlich J. 1993, {\it Gravity in
Non-Commutative Geometry}, Commun. Math. Phys. {\bf 155} 205.

Connes A. 1986, {\it Non-Commutative Differential Geometry}, Publications
of the Inst. des Hautes Etudes Scientifique. {\bf 62} 257.

Connes A., Lott J. 1990, {\it Particle Models and Noncommutative Geometry},
in `Recent Advances in Field Theory', Nucl. Phys. Proc. Suppl. {\bf B18} 29.

--- 1991, {\it The metric aspect of non-commutative geometry},
Proceedings of the Carg\`ese Summer School, (to appear).

Dubois-Violette M., Kerner R., Madore J. 1989a, {\it Gauge bosons in a
noncommutative geometry}, Phys. Lett. {\bf B217} 485.

--- 1989b {\it Classical bosons in a noncommutative geometry}, Class.
Quant. Grav. {\bf 6} 1709.

Dubois-Violette M., Madore J., Masson T., Mourad J. 1994, {\it Linear
Connections on the Quantum Plane}, Preprint LPTHE Orsay 94/94

Dubois-Violette M., Michor P. 1994, {\it D\'erivations et calcul
diff\'erentiel non-commuta\-tif II}, C. R. Acad. Sci. Paris {\bf 319}
S\'erie I 927.

Klim\v c\'ik C., Pompo\v s A., Sou\v cek. V. 1994, {\it Grading of
Spinor Bundles and Gravitating Matter in Noncommutative Geometry}, Lett.
Math. Phys.  {\bf 30}, 259.

Koszul J.L. 1960, {\it Lectures on Fibre Bundles and Differential Geometry},
Tata Institute of Fundamental Research, Bombay.

Landi G., Nguyen Ai Viet, Wali K.C. 1994, {\it Gravity and
electromagnetism in noncommutative geometry}, Phys. Lett. {\bf B326} 45.

Madore J. 1990, {\it Modification of Kaluza-Klein Theory}, Phys. Rev.
{\bf D41} 3709.

Madore J., Masson T., Mourad J. 1994, {\it Linear Connections on Matrix
Geometries}, Preprint LPTHE Orsay 94/96.

Madore J., Mourad J. 1993, {\it Algebraic-Kaluza-Klein Cosmology}, Class.
Quant. Grav. {\bf 10} 2157.

Mourad. J. 1994, {\it Linear Connections in Non-Commutative Geometry},
Class. Quant. Grav. {\bf } (to appear).

Sitarz A. 1994, {\it Gravity from Noncommutative Geometry}, Class.
Quant. Grav. {\bf 11} 2127.

Sitarz A. 1995, {\it On some aspects of linear connections in
noncommutative geometry}, Jagiellonian University Preprint.

\bye